# Hierarchical organization of chiral rafts in colloidal membranes

Prerna Sharma[1,#], Andrew Ward[1,#], T. Gibaud[2], Michael F. Hagan[1] and Zvonimir Dogic[1]

[1]Department of Physics, Brandeis University, 415 South St., Waltham, MA 02454, USA
[2]Laboratoire de Physique, Ecole Normale Superieure de Lyon, Universit de Lyon I, CNRS/UMR 5672, 46 alle d'Italie, Lyon 69007, France
* Correspondence to: zdogic@brandeis.edu
[#]These authors contributed equally.

Liquid-liquid phase separation is ubiquitous in suspensions of nanoparticles, proteins and colloids. It plays an important role in gel formation, protein crystallization and perhaps even as an organizing principle in cellular biology[1,2]. With a few notable exceptions[3,4], surface-tension-minimizing liquid droplets in bulk suspensions continuously coalesce, increasing in size without bound until achieving macroscale phase separation. In comparison, the phase behavior of colloids, nanoparticles or proteins confined to interfaces, surfaces or membranes is significantly more complex[5-11]. Inclusions distort the local interface structure leading to interactions that are fundamentally different from the well-studied interactions mediated by isotropic solvents[12,13]. Here, we investigate liquid-liquid phase separation in monolayer membranes composed of dissimilar chiral colloidal rods. We demonstrate that colloidal rafts are a ubiquitous feature of binary colloidal membranes. We measure the raft free energy landscape by visualizing its assembly kinetics. Subsequently, we quantify repulsive raft-raft interactions and relate them to directly imaged raft-induced membrane distortions, demonstrating that particle chirality plays a key role in this microphase separation. At high densities, rafts assemble into cluster crystals which constantly exchange monomeric rods with the background reservoir to maintain a self-limited size. Lastly, we **demonstrate that rafts can form bonds to assemble into higher-order suprastructures. Our work demonstrates that membrane-mediated liquid-liquid phase separation can be fundamentally different from the well-characterized behavior of bulk liquids. It outlines a robust membrane-based pathway for assembly of monodisperse liquid clusters which is complementary to existing methods which take place in bulk suspensions[14-16]. Finally, it reveals that chiral inclusions in membranes acquire long-ranged repulsive interactions, which might play a role in stabilizing assemblages of finite size[11,17].**

In the presence of non-adsorbing polymer, mono-disperse rod-like viruses experience effective depletion attractions that drive their lateral association. These interactions can lead to assembly of one-rod-length-thick colloidal monolayer membranes that are held together by the osmotic pressure of the enveloping polymer suspension[18]. We allow membranes to sediment to the bottom of the sample chambers in which case the constituent rods point in the $z$ direction, while all images are taken in the $x$-$y$ plane. Although they differ on molecular scales, the long-wavelength fluctuations of colloidal monolayers and lipid bilayers are described by the same free energy. In this work, we investigated the behavior of colloidal membranes containing a mixture of two rods: 880 nm long rod-like *fd*-Y21M virus and 1200 nm long M13KO7 virus[19]. Membranes were prepared by adding a depletant to a dilute isotropic *fd*-Y21M/M13KO7 mixture. For all parameters investigated, both rods co-assembled into binary membranes. At low depletant concentration the rods remained homogeneously mixed throughout the membrane (Fig. 1b,d), while at high depletant concentration they bulk phase separated (Fig. 1b,f). Surprisingly, at intermediate concentrations we observed the formation of highly-monodisperse micron-sized droplets (colloidal rafts) enriched in short rods and floating in the background of long rods (Fig. 1e). Colloidal rafts did not coarsen with time suggesting that they are equilibrium structures.

To test this hypothesis, we brought two membranes, each containing only one type of virus, into close proximity (Fig. 1a). Upon coalescence, the interface separating dissimilar rods became unstable as short rods invaginated into the long-rod fluid, giving rise to an interconnected network of rafts held together by thin metastable bridges (Fig. 1c, Supplementary Movie 1). Over time the conjoining liquid bridges disintegrated, leading to the formation of isolated finite-sized rafts thus demonstrating that colloidal rafts have a lower free-energy than a bulk separated phase. We have explored the behavior of six phase-separating mixtures comprised of viruses of varying physical properties. In each case, we observed micro-separated phases at intermediate polymer concentrations, suggesting that rafts are a ubiquitous feature of colloidal membranes. Here, we investigate the behavior of the *fd*-Y21M/M13KO7 mixture, utilizing the unique features of colloidal membranes that allow us to optically manipulate rafts, quantify their nucleation dynamics as well as track the dynamics of the constituent rods.

We determined the raft free-energy landscape by measuring how $k_{on}$ and $k_{off}$, the rates at which rods associate and dissociate from a raft, depend on the raft size. We measured $k_{off}$ using Fluorescence Recovery After Photobleaching (FRAP)[20]. Using a focused beam, we bleached an isolated raft of predetermined size. The fluorescence signal recovered within minutes, indicating exchange of bleached raft-bound rods with unbleached background rods (Fig. 2a, Supplementary Movie 2). The fluorescence recovery curves were fit by single exponentials (Fig. 2e). If the raft size does not change during of experiment, the fluorescence recovery time constant is $1/k_{off}$ (Methods). By performing FRAP measurements on rafts of varying radii we determined the dependence of $k_{off}$ on raft size (Fig. 2h).

To measure $k_{on}$, we created a raft population with heterogeneous radii R and quantified their subsequent growth rates (Fig. 2b, c, f, Supplementary Movie 3). Rafts with $R<R_{critical}$ quickly evaporated into the background membrane (Fig. 2g). When $R_{equilibrium}>R>R_{critical}$ rafts grew until attaining the equilibrium size. Finally rafts with $R>R_{equilibrium}$ shrank slowly until achieving equilibrium. The evolution of raft size is governed by the kinetic equation:

$$\frac{dN(t)}{dt} = -k_{off}N(t) + k_{on}C_{BG},$$

where $N(t)$ is the number of raft-bound rods and $C_{BG}$ is the concentration of short fluorescently-labeled rods in the background. $dN(t)/dt$ is determined from size evolution experiments (Fig. 2g), $k_{off}$ is determined through FRAP experiments, and $C_{BG}$ is proportional to mean fluorescence of the background membrane. The only unknown is $k_{on}$, making it possible to determine its dependence on raft size. We found that $k_{on}$ increases weakly with increasing raft size, consistent with 2D diffusion-limited association kinetics. Having measured both $k_{on}$ and $k_{off}$ (Fig. 2h), we calculate $\Delta G(N) = k_B T \left(\frac{k_{off}NA_{rod}}{k_{on}}\right)$, where $NA_{rod}$ is the raft area and $\Delta G(N)$ is the energy cost of adding a single particle to a raft (Fig. 2i). The minimum of the excess free energy satisfies $\Delta G - k_B T \ln(C_{BG}A_{rod})=0$, and determines the equilibrium raft size, which corresponds to ~24000 rods. Integrating $\Delta G$ over $N$ yields the total free-energy $G(N)$. The width of the parabola determines the fluctuations in raft area, which is ~1400 particles, corresponding to 30 nm radius fluctuations. Such fluctuations are below the optical microscope resolution limit; consistent with our observation of a monodisperse raft distribution in equilibrium.

To independently validate our analysis, we performed single-molecule lifetime experiments by tracking a low-fraction of fluorescently labeled short rods in an otherwise unlabeled membrane (Fig. 2d, Supplementary Movie 4). Overlaying fluorescent images onto simultaneously acquired phase contrast images revealed raft-bound rods as a function of time, $N_f(t)$ (Fig. 2j). From such data we find that for equilibrium rafts $k_{on}$=3.7 min$^{-1}$μm$^2$ and $k_{off}$=0.15 min$^{-1}$, which is in agreement with FRAP measurements (Methods). To summarize colloidal membranes enabled us to directly visualize raft nucleation dynamics and measure their free-energy landscape. So far such measurements were possible in a very few experimental systems[14,21].

With an understanding of single-raft assembly kinetics, we subsequently measured membrane-mediated raft-raft interactions by adapting a blinking optical tweezers technique[22]. To manipulate rafts, which are repelled from a trapping laser, we created an optical plow composed of multiple beams. Using two plows we pushed a raft pair into close proximity. Subsequently, we switched off the trap and quantified raft trajectories as they were pushed apart by repulsive interactions (Supplementary Movie 5). Repeating this procedure dozens of times yielded the inter-raft potential (Methods). The effective repulsive potential is described by an exponential with a characteristic length scale of 0.65μm (Fig. 3a). The measured raft-raft interactions can be quantitatively related to raft-induced membrane distortions visualized by LC-PolScope. This technique yields images whose intensity indicates the local retardance, which is related to the local tilt of rods away from the membrane normal vector (Fig. 3c)[23]. The interior of a homogeneous one-component membrane viewed from above lacks optical anisotropy and appears dark in the LC-PolScope. In contrast, rafts imaged with LC-PolScope exhibited a spatially varying retardance, indicating local twisting of raft-bound rods.

Such deformations can be understood by considering that *fd* Y21M and M13KO7 have right- and left-handed chirality, respectively[19,24]. As a consequence, the interaction energy between membrane-embedded viruses is minimized when neighboring rods are not parallel to each other, but are twisted at a small but finite angle. The radially averaged retardance profile reveals that, starting from the raft center, *fd* Y21M twist with a right-handed sense, leading to increasing rod tilt away from the monolayer normal. The tilt attains maximum at the raft's edge. Moving past the edge, the membrane is enriched in left-handed M13KO7. The left-handed twist induces tilting of rods back toward the membrane normal. Larger rafts attain a larger maximum edge tilt; however, the characteristic length over which twist penetrates into the background membrane is independent of raft size (Fig. 3d,e). The length scales associated with raft repulsion and raft induced twist deformation are comparable, suggesting a possible link.

Homo-chiral colloidal membranes are inherently frustrated, since the constituent rods cannot simultaneously twist locally and assemble into a monolayer globally. Consequently, twist is expelled from the membrane interior to the edges. Increasing the rod chirality raises the free energy of interior untwisted rods while lowering the free energy of edge-bound twisted rods, leading to chiral control of edge line tension[25,26]. For uniformly mixed binary membranes, the same constraints enforce untwisting of all constituent chiral rods, thus raising their free energy. However, formation of right-handed rafts in the left-handed background allows twist to penetrate into the membrane interior, lowering raft free energy. Besides explaining raft stability, this picture also explains the origin of inter-raft repulsion. Bringing two rafts together closer than the twist penetration length requires energetically costly untwisting of chiral rods. Our hypothesis predicts that the range of raft repulsion is independent of cluster size, while its strength increases linearly with cluster size. Measured potentials between clusters of different sizes agree with these predictions (Fig. 3a, b). Furthermore, when the magnitude of the repulsive interactions is rescaled by the maximum twist at the rafts edge, all potentials collapse on top of each other, thus establishing a quantitative relationship between raft-induced chiral distortions and membrane-mediated raft interactions. These experiments describe a chiral mechanism for formation of finite-sized rafts that is complementary to others proposed mechanisms[27-29].

The emergent raft repulsions can be used to assemble higher-order structures. To explore this capability, we have increased raft densities and measured the raft radial distribution function, $g(r)$, of resulting assemblages (Supplementary Movie 6). In the dilute limit, $g(r)$ is zero for small separations, indicative of strong repulsive interactions, and constant elsewhere (Fig. 4a). At intermediate raft densities, $g(r)$ exhibits liquid-like oscillations that decay after a few coordination shells (Fig. 4b). At the highest densities, rafts form a 2D crystal-like structure as evidenced by sharp peaks in $g(r)$ (Fig. 4c). This behavior is reminiscent of conventional 2D repulsive colloids. However, there are also important differences, since colloidal crystals are assembled from immutable solid particles while the membrane-embedded crystals are assembled from highly adaptable, size-adjustable rafts, each of which constantly exchanges rods with the background. Moreover, while conventional droplets in bulk suspensions always assume a surface-minimizing spherical shape, the behavior of rafts embedded in a membrane is significantly more complex. For example, it is possible to assemble supra-rafts, which are liquid droplets with distinctly non-spherical shapes. At higher ionic strength and/or depletant concentrations, isolated rafts bind together to form dimers, trimers and higher-order rafts bound together by thin liquid bridges (Fig. 4d, Supplementary Movie 7). Using such bonding rules it is possible to assemble structures of almost arbitrary complexity such as a classical bead-spring polymer (Fig. 4e, Supplementary Movie 8).

In conclusions, we have demonstrated assembly of thermodynamically stable, monodisperse, surfactant-free liquid droplets which are driven by membrane-mediated repulsive chiral interactions that are orders of magnitude longer-ranged than attractive depletion interactions[30]. These results may be relevant for biological membranes, which contain diverse chiral inclusions. In particular they suggest chirality as a plausible mechanism that stabilizes assembly of finite-sized clusters such as lipid rafts. Intriguingly, it is known that cholesterol, an effective chiral dopant of conventional liquid crystals, is essential for the assembly of lipid rafts.

**Acknowledgments:** We acknowledge useful discussion with R. B. Meyer. This work was supported by the US National Science Foundation (NSF-MRSEC-0820492, NSF-DMR-0955776, NSF-MRI-0923057) and the Petroleum Research Fund (ACS-PRF 50558-DNI7). We acknowledge use of the Brandeis MRSEC optical microscopy facility.

**Author Contributions**: P.S., A.W. and Z.D. conceived and designed the experiments. P.S and A.W. performed the experiments. M.F.H. designed the theoretical models and P.S., A.W., Z.D. and M.F.H. interpreted the experiments. T.G. provided material samples. P.S. and Z.D. wrote the manuscript.

**Author Information:** Reprints and permissions information is available at www.nature.com/reprints. The authors declare no competing financial interests. Correspondence and request for materials should be addressed to Z. D. (zdogic@brandeis.edu)

**References**

1. tenWolde, P. R. & Frenkel, D. Enhancement of protein crystal nucleation by critical density fluctuations. *Science* **277**, 1975-1978 (1997).
2. Hyman, A. A. & Simons, K. Beyond Oil and Water-Phase Transitions in Cells. *Science* **337**, 1047-1049, doi:DOI 10.1126/science.1223728 (2012).
3. Stradner, A. *et al.* Equilibrium cluster formation in concentrated protein solutions and colloids. *Nature* **432**, 492-495, doi:Doi 10.1038/Nature03109 (2004).
4. Groenewold, J. & Kegel, W. K. Anomalously large equilibrium clusters of colloids. *J Phys Chem B* **105**, 11702-11709, doi:Doi 10.1021/Jp011646w (2001).
5. Weis, R. M. & Mcconnell, H. M. Two-Dimensional Chiral Crystals of Phospholipid. *Nature* **310**, 47-49, doi:Doi 10.1038/310047a0 (1984).
6. Dietrich, C. *et al.* Lipid rafts reconstituted in model membranes. *Biophys J* **80**, 1417-1428 (2001).
7. Dinsmore, A. D. *et al.* Colloidosomes: Selectively permeable capsules composed of colloidal particles. *Science* **298**, 1006-1009, doi:DOI 10.1126/science.1074868 (2002).
8. Lin, Y., Skaff, H., Emrick, T., Dinsmore, A. D. & Russell, T. P. Nanoparticle assembly and transport at liquid-liquid interfaces. *Science* **299**, 226-229, doi:DOI 10.1126/science.1078616 (2003).
9. Veatch, S. L. & Keller, S. L. Separation of liquid phases in giant vesicles of ternary mixtures of phospholipids and cholesterol. *Biophys J* **85**, 3074-3083 (2003).
10. Baumgart, T., Hess, S. T. & Webb, W. W. Imaging coexisting fluid domains in biomembrane models coupling curvature and line tension. *Nature* **425**, 821-824, doi:Doi 10.1038/Nature02013 (2003).
11. Lingwood, D. & Simons, K. Lipid Rafts As a Membrane-Organizing Principle. *Science* **327**, 46-50, doi:DOI 10.1126/science.1174621 (2010).

12  Goulian, M., Bruinsma, R. & Pincus, P. Long-Range Forces in Heterogeneous Fluid Membranes (Vol 22, Pg 145, 1993). *Europhys Lett* **23**, 155-155, doi:Doi 10.1209/0295-5075/23/2/014 (1993).
13  Dan, N., Berman, A., Pincus, P. & Safran, S. A. Membrane-Induced Interactions between Inclusions. *J Phys Ii* **4**, 1713-1725 (1994).
14  Meng, G. N., Arkus, N., Brenner, M. P. & Manoharan, V. N. The Free-Energy Landscape of Clusters of Attractive Hard Spheres. *Science* **327**, 560-563, doi:10.1126/science.1181263 (2010).
15  Chen, Q. *et al.* Supracolloidal Reaction Kinetics of Janus Spheres. *Science* **331**, 199-202, doi:DOI 10.1126/science.1197451 (2011).
16  Wang, Y. F. *et al.* Colloids with valence and specific directional bonding. *Nature* **491**, 51-U61, doi:Doi 10.1038/Nature11564 (2012).
17  Sarasij, R. C., Mayor, S. & Rao, M. Chirality-induced budding: A raft-mediated mechanism for endocytosis and morphology of caveolae? *Biophys J* **92**, 3140-3158, doi:DOI 10.1529/biophysj.106.085662 (2007).
18  Barry, E. & Dogic, Z. Entropy driven self-assembly of nonamphiphilic colloidal membranes. *P Natl Acad Sci USA* **107**, 10348-10353, doi:DOI 10.1073/pnas.1000406107 (2010).
19  Barry, E., Beller, D. & Dogic, Z. A model liquid crystalline system based on rodlike viruses with variable chirality and persistence length. *Soft Matter* **5**, 2563-2570, doi:Doi 10.1039/B822478a (2009).
20  Sprague, B. L., Pego, R. L., Stavreva, D. A. & McNally, J. G. Analysis of binding reactions by fluorescence recovery after photobleaching. *Biophys J* **86**, 3473-3495, doi:10.1529/biophysj.103.026765 (2004).
21  Gasser, U., Weeks, E. R., Schofield, A., Pusey, P. N. & Weitz, D. A. Real-space imaging of nucleation and growth in colloidal crystallization. *Science* **292**, 258-262, doi:DOI 10.1126/science.1058457 (2001).
22  Crocker, J. C. & Grier, D. G. Microscopic Measurement of the Pair Interaction Potential of Charge-Stabilized Colloid. *Phys Rev Lett* **73**, 352-355, doi:DOI 10.1103/PhysRevLett.73.352 (1994).
23  Barry, E., Dogic, Z., Meyer, R. B., Pelcovits, R. A. & Oldenbourg, R. Direct Measurement of the Twist Penetration Length in a Single Smectic A Layer of Colloidal Virus Particles. *J Phys Chem B* **113**, 3910-3913, doi:Doi 10.1021/Jp8067377 (2009).
24  Tombolato, F., Ferrarini, A. & Grelet, E. Chiral nematic phase of suspensions of rodlike viruses: Left-handed phase helicity from a right-handed molecular helix. *Phys Rev Lett* **96**, doi:10.1103/PhysRevLett.96.258302 (2006).
25  Gibaud, T. *et al.* Reconfigurable self-assembly through chiral control of interfacial tension. *Nature* **481**, 348-U127, doi:Doi 10.1038/Nature10769 (2012).
26  Kaplan, C. N. & Meyer, R. B. Colloidal membranes of hard rods: unified theory of free edge structure and twist walls. *Soft Matter* **10**, 4700-4710 (2014).
27  Rozovsky, S., Kaizuka, Y. & Groves, J. T. Formation and spatio-temporal evolution of periodic structures in lipid bilayers. *J Am Chem Soc* **127**, 36-37, doi:Doi 10.1021/Ja046300o (2005).

28      Reynwar, B. J. *et al.* Aggregation and vesiculation of membrane proteins by curvature-mediated interactions. *Nature* **447**, 461-464, doi:Doi 10.1038/Nature05840 (2007).
29      Ursell, T. S., Klug, W. S. & Phillips, R. Morphology and interaction between lipid domains. *P Natl Acad Sci USA* **106**, 13301-13306, doi:DOI 10.1073/pnas.0903825106 (2009).
30      Seul, M. & Andelman, D. Domain Shapes and Patterns - the Phenomenology of Modulated Phases. *Science* **267**, 476-483, doi:DOI 10.1126/science.267.5197.476 (1995).
31      Dogic, Z. & Fraden, S. Development of model colloidal liquid crystals and the kinetics of the isotropic-smectic transition. *Philosophical Transactions of the Royal Society a-Mathematical Physical and Engineering Sciences* **359**, 997-1014 (2001).
32      Maniatis, T., Sambrook, J., Fritsch, E. *Molecular Cloning*.  (1989).
33      Lettinga, M. P., Barry, E. & Dogic, Z. Self-diffusion of rod-like viruses in the nematic phase. *Europhys Lett* **71**, 692-698, doi:10.1209/epl/i2005-10127-x (2005).
34      Lau, A. W. C., Prasad, A. & Dogic, Z. Condensation of isolated semi-flexible filaments driven by depletion interactions. *Epl* **87**, doi:10.1209/0295-5075/87/48006 (2009).
35      Oldenbourg, R. & Mei, G. New Polarized-Light Microscope with Precision Universal Compensator. *J Microsc-Oxford* **180**, 140-147 (1995).
36      Zakhary, M. J. *et al.* Imprintable membranes from incomplete chiral coalescence. *Nat. Commun.* **5**, 9, doi:10.1038/ncomms4063 (2014).
37      Gardiner, C. W. *Handbook of Stochastic methods.*,  (Springer-Verlag, 1985).
38      Crocker, J. C. & Grier, D. G. Methods of digital video microscopy for colloidal studies. *J Colloid Interf Sci* **179**, 298-310, doi:10.1006/jcis.1996.0217 (1996).

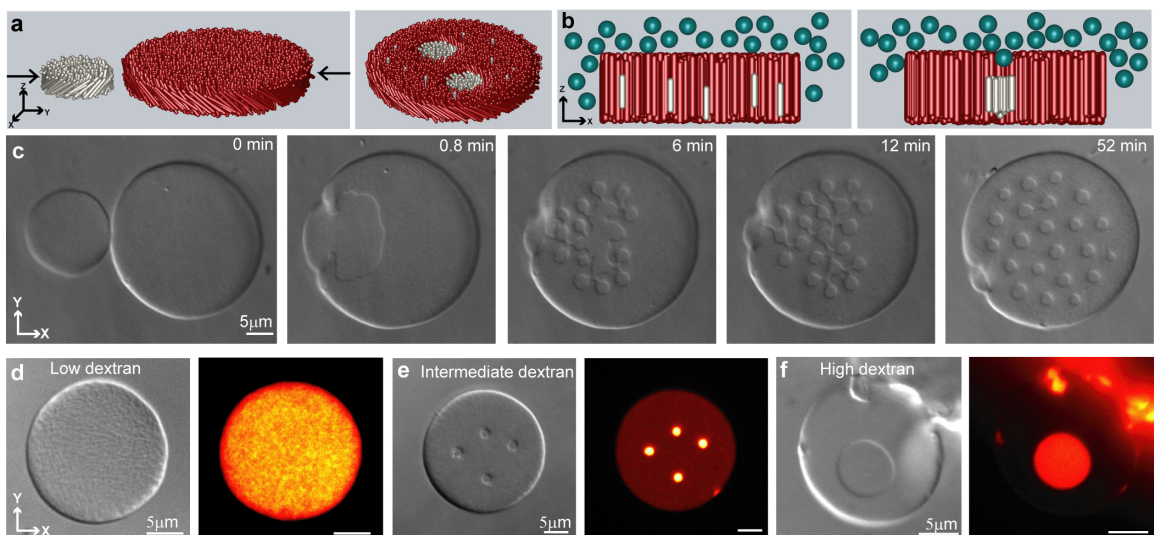

**Fig.1. Membrane-mediated assembly of monodisperse chiral colloidal rafts. a.** Coalescence of membranes comprised of short and long rods with opposite chirality. **b.** Clustering of short rods dissolved in a long-rod membrane is driven by excluded volume interactions. **c.** Coalescence of right-handed *fd*-Y21m and left-handed M13KO7 membranes leads to the formation of finite sized rafts. **d.** DIC and fluorescence images of a*fd*-Y21M/MK13KO7 membrane at 34 mg/ml Dextran concentration show homogeneous mixing. *fd*-Y21M are fluorescently labeled. **e.** Formation of finite-sized clusters enriched in fluorescently labeled *fd*-Y21M at intermediate dextran concentrations (38 mg/ml). The image also indicates that a fraction of shorter fluorescent *fd*-Y21M rods are dissolved in the background membrane. **f.** Bulk phase separation is observed at the highest dextran concentrations (52 mg/ml).

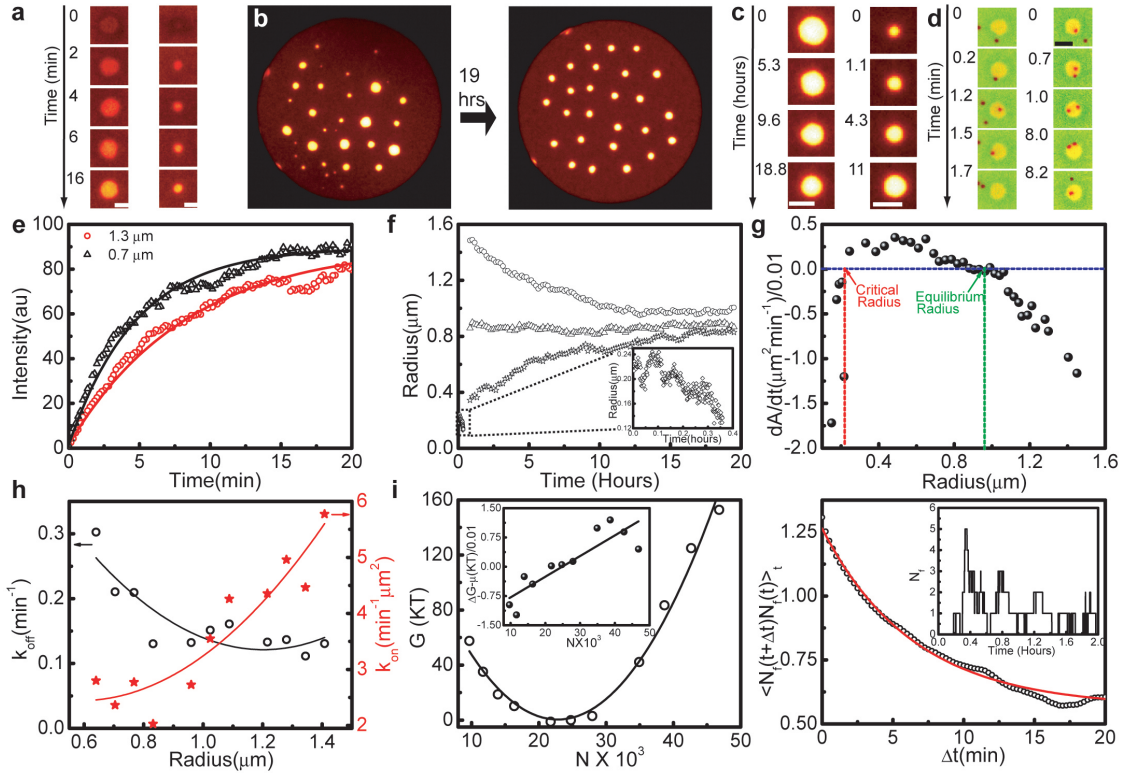

**Fig. 2. Single-molecule analysis reveals raft nucleation dynamics and assembly kinetics. a.** The fluorescence of a bleached cluster recovers after a few minutes, revealing kinetics of rod exchange between the raft and the background. **b.** Initially polydisperse rafts equilibrate in size after many hours. **c.** Expansion of an undersized raft and contraction of an oversized raft. **d.** Fluorescently labeled *fd*-Y21M rods associate and dissociate from a raft, revealing kinetics at the single molecule level. **e.** Exponential recovery of cluster fluorescence intensity after a photobleaching event yields $k_{off}$ for rafts with different sizes. **f.** Time evolution of cluster expansion depends on the initial cluster size. Inset: A sub-critical raft quickly evaporates into the membrane background. **g.** The dependence of raft expansion rates on raft size directly reveals the critical nucleus size and equilibrium size. **h.** Combining single-raft fluorescence recovery measurements with raft expansion rates reveals dependence of $k_{off}$ (circles) and $k_{on}$ (stars) on raft size. **i.** The raft free energy landscape exhibits a minimum indicating equilibrium rafts. Inset: *ΔG*, the free energy change associated with adding a single virus to a colloidal raft. **j.** Exponential decay of the autocorrelation function of the number of raft-bound rods, *N(t)*, yields both $k_{off}$ and $k_{on}$. For equilibrium-sized rafts single-molecule analysis agrees with measurements in panel h. Inset: Plot of *N(t)*. All scale bars 2.5 μm.

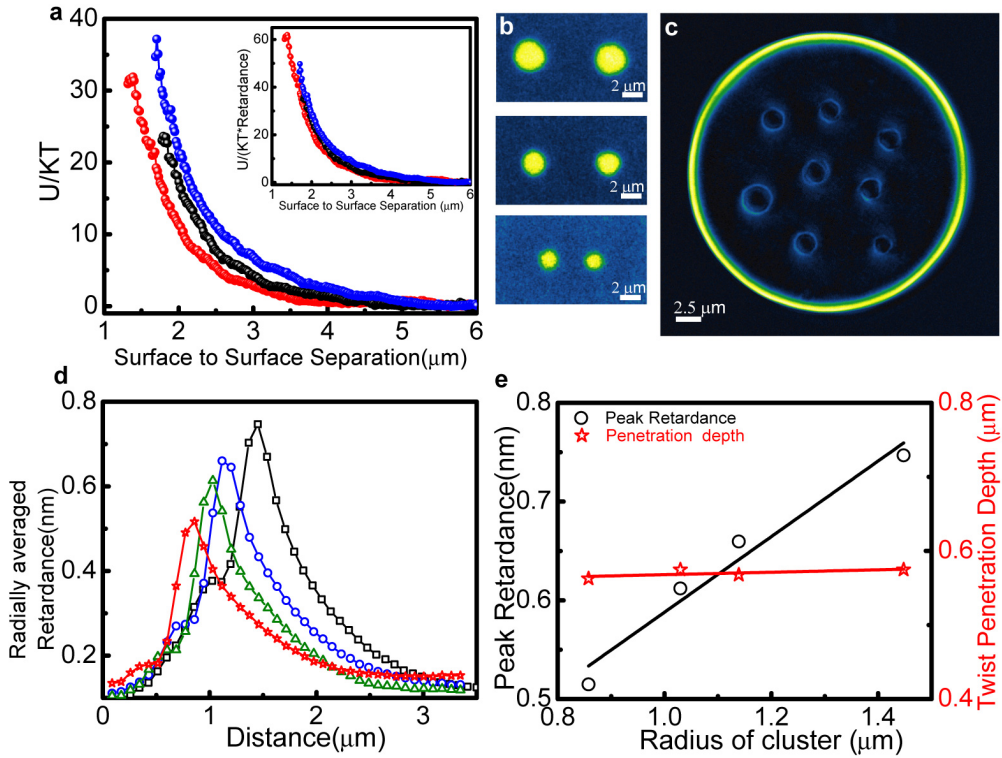

**Fig. 3. Raft-induced chiral twist governs membrane-mediated raft repulsions. a.** Effective pair interaction potential obtained using the blinking trap technique for clusters with diameters of 2.9 μm (blue), 2.1 μm (black), and 1.6 μm (red). Inset: Interaction potential rescaled by maximum tilt angle. **b.** Images of cluster pairs for which interactions were measured. **c.** LC-PolScope image of heterogeneous rafts embedded in a membrane. **d.** Radially averaged retardance profile for rafts of increasing size. **e.** The peak retardance (maximum tilt angle) increases linearly with increasing raft size. The twist penetration depth is independent of domain size.

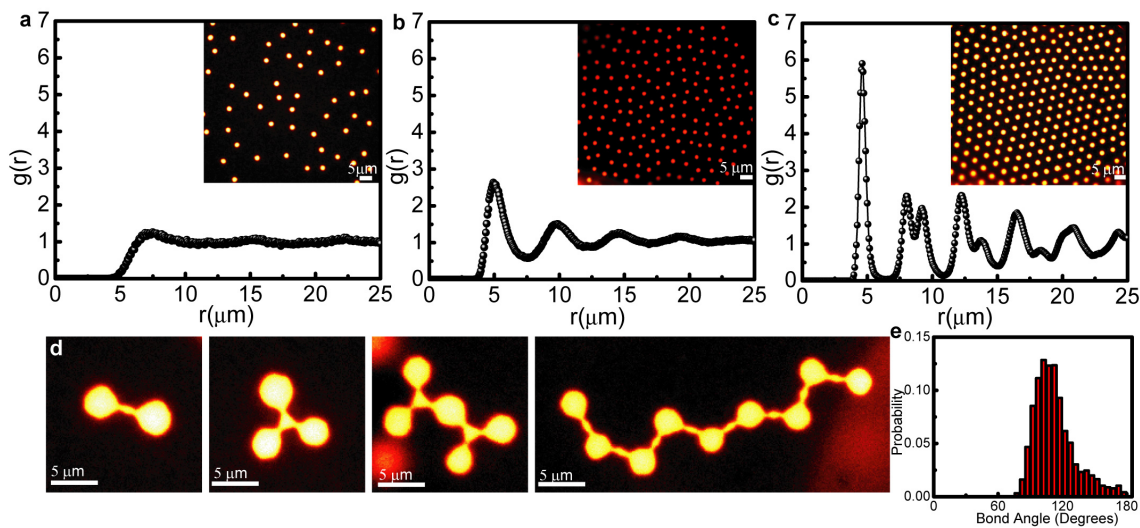

**Fig. 4. Assembly of raft crystals and aspherical supra-rafts. a.** Radial distribution function *g(r)* of dilute membrane-embedded clusters. The M13KO7/*fd*-Y21M stoichiometric ratio is 6:1. Inset: A representative image of the membrane. **b.** At intermediate raft densities *g(r)* exhibits a liquid-like structure (M13K07/*fd*-Y21M ratio 4:1). **c.** At the highest densities rafts organize into a 2D cluster crystal (M13K07/*fd*-Y21M ratio 2.2:1). **d.** Rafts can be permanently bonded by thin liquid bridges to form complex architectures. Fluorescence images of a raft dimer, trimer, pentamer and polymer-like nonamer. **e.** Probability distribution of bond angles in the bead-spring nonamer.

**Materials and Methods**

**Virus growth and sample preparation**: Colloidal rafts were assembled from membranes containing a mixture of filamentous bacteriophages *fd*-Y21M and M13K07. The rod-like *fd*-Y21M virus has a 6 nm diameter, 880 nm contour length and a 9.9 μm persistence length[19]. M13KO7 has the same diameter, but a 1200 nm contour length and persistence length of 2.8 μm[31]. Both systems exhibit isotropic, cholesteric and smectic phases with increasing virus concentration. However, *fd*-Y21M forms a cholesteric with a right-handed twist while the M13K07 cholesteric phase twists in the opposite direction[19,24].

M13K07 and *fd*-Y21M viruses were purified using standard biological protocols[32]. To remove end-to-end dimers or longer viruses that destabilize colloidal membranes, purified virus suspensions were fractionated trough the isotropic-nematic phase transition[20]. Only the isotropic fraction, enriched in nominal length viruses, was used for assembling colloidal membranes. Viruses were suspended in Tris-HCl buffer (20 mM, pH=8.0) to which 100 mMNaCl was added to screen electrostatic interactions. For fluorescence visualization, primary amines of the major coat protein of *fd*-Y21M were labeled with amine reactive fluorophore (DyLight-NHS ester 550, Thermo Fisher)[33]. There are about 2700 labeling sites available on the virus surface. However, each virus was labeled at a low volume fraction (~30 dye molecules per virus), ensuring that the labeling did not affect the system phase behavior. *fd*-Y21M and M13K07 were mixed at known stoichiometric ratios. The non-adsorbing polymer Dextran (MW 500,000, Sigma-Aldrich) was added to this suspension and the resultant suspension was injected into an observation chamber. The chamber was made from a microscope slide and a coverslip (Goldseal, Fisher Scientific) separated by a spacer consisting of a non-stretched Parafilmand sealed using ultraviolet-cured glue (Norland Optical). Glass surfaces were thoroughly cleaned with a hot 1% soap solution (Hellmanex, Hellma Analytics). Samples prepared in such a manner remained stable for weeks or even months. The final concentration of bi-disperse virus mixture was 2.5 mg/ml. All the measurements were performed for membranes assembled at 40 mg/ml Dextran concentration. Membranes initially assemble in the bulk suspension but due to their density they eventually sediment onto the coverslip. To prevent non-specific absorption, glass surfaces were coated with a poly-acrylamide brush which suppresses the depletion interaction between viruses and the glass walls[34].

**Optical Microscopy Methods:** Within a liquid-like colloidal monolayer membrane, rods with opposite chirality phase separated to form finite-sized colloidal rafts, with the exact behavior depending on the depletant concentration. Raft structures and dynamics were examined with a number of complementary optical microscopy techniques. DIC, phase contrast and fluorescence images were acquired using an inverted microscope (Nikon TE2000) equipped with an oil-immersion objective (1.3 NA, 100X Plan-Fluor) and connected to charge-coupled-device (CCD) camera (Andor Clara). Fluorescently labeled *fd*-Y21M virus was imaged using the same setup equipped with Mercury-Halide epi-fluorescence source and a rhodamine filter cube (excitation wavelength 532 – 554 nm, emission wavelength 570 – 613 nm). The exposure time for acquisition of fluorescent images was 100 ms.

Spatial variations of rod tilt with respect to the membrane normal were determined using a quantitative polarized light microscope (LC-Polscope, Cambridge Research and Instrumentation)[35]. LC-PolScope yields images in which the intensity of each pixel is proportional to local sample birefringence. Subtraction of the background birefringence enables visualization of sub-nanometer birefringent images with LC-PolScope, something that is not observable with conventional techniques. When a membrane lies in the image plane, it is possible to quantitatively translate LC-PolScope birefringence signal into local tilting of the rods away from the membrane normal[23,36].

**Determination of $k_{off}$ using single-raft FRAP:** DyLight 550 attached to *fd*-Y21M surface has an excitation wavelength of 562nm. An isolated raft of predetermined size enriched with fluorescently labeled *fd*-Y21M was illuminated with a focused laser beam (~100 mW, 514nm). Under these conditions, raft-bound rods were photo-bleached instantaneously for all practical purposes. The exchange of rods between rafts and the background membrane causes fluorescence to recover on time scale of minutes. This process can be modeled using kinetic equations. $N_F(t)$ and $N_B(t)$ respectively describe the number of fluorescent and bleached rods in a raft at time *t*. Assuming that the raft size does not change on fluorescence recovery timescales and that the concentration of bleached rods in the membrane background is negligible, the time evolution of $N_F(t)$ and $N_B(t)$ is given by:

$$\frac{dN_f(t)}{dt} = -k_{off} N_f(t) + k_{on} C_{BG}$$

$$\frac{dN_B(t)}{dt} = -k_{off} N_B(t)$$

where $k_{off}$ and $k_{on}$ are rod dissociation and association rates and $C_{BG}$ is the concentration of fluorescent *fd*-Y21M rods in the background membrane[20]. The experimentally measured total raft intensity is given by $I_{tot}(t)=\alpha N_F(t)+\beta N_B(t)$ where $\alpha$ and $\beta$ are the intensities of single fluorescent labeled and bleached viruses, respectively. Perfect bleaching would result in $\beta=0$. However, we find that this not the case as bleached rods retain finite fluorescence. Solving for $I_{tot}(t)$ yields:

$$I_{tot}(t) = \exp(-k_{off}t)(I_0 - I_{avg}^{BG}\frac{k_{on}}{k_{off}}) + I_{avg}^{BG}\frac{k_{on}}{k_{off}}, \qquad (1)$$

where $I_0$ is the raft fluorescence intensity after completion of the photo-bleaching event and $I_{avg}^{BG}$ is the mean fluorescence intensity of the background membrane. Eq. (1) shows that the fluorescence recovery process can be approximated by a single exponential rate, in agreement with experimental findings. The exponential time constant yields the rod dissociation constant $1/k_{off}$ and its dependence on raft size (Fig. 2e,h).

**Evolution of polydisperse rafts yields $k_{on}$:** Heterogeneously sized rafts evolve toward equilibrium, a state characterized by rafts of uniform size. Using optical tweezers we fragmented/fused a small fraction of rafts, thus creating an artificially polydisperse population, and then quantified their subsequent size-evolution. The long time-lapse movies required for this measurement were acquired on a microscope equipped with auto-focus capabilities (Perfect-Focus, Nikon Ti-E). The rate equation governing the evolution of raft size is:

$$\frac{dN(t)}{dt} = -k_{off}N(t) + k_{on}C_{BG}, \qquad (2)$$

where *N(t)* is the total number of raft bound rods at time *t*. Cluster size evolution experiments yielded $\frac{dN(t)}{dt}$ and *N(t)* (Fig. 2g), while $C_{BG}$ was extracted from fluorescent membrane images, and $k_{off}$ and its size were obtained from previously described single raft FRAP measurements. It follows that $k_{on}$ and its dependence on raft size can be uniquely determined from Eq. 2, as all other parameters are determined from independent experiments.

**Measuring the raft free energy landscape:** The change in free energy, *ΔG*, to add a single rod to a raft containing *N* rods is given by:

$$\frac{k_{on}}{k_{off}} = A_{rod} N e^{-\Delta G / k_B T}, \qquad (3)$$

Where $A_{rod}$ is the effective area occupied by a single *fd*-Y21M virus, *N* is the number of rods in a raft and $A_{rod}N$ is the raft size. Assuming dilute suspension conditions, the chemical potential of *fd*-Y21M rods dissolved in the background membrane is given by: $\mu = K_B T \ln(C_{BG} A_{rod})$. The net free energy change associated with taking a single *fd*-Y21M virus from the background and inserting it into a raft containing *N* rods is given by *ΔG-μ* (inset Fig. 2i). The absolute raft free energy landscape up to a constant offset was obtained by numerically integrating *ΔG-μ* using a cumulative trapezoidal rule (Fig. 2i).

**Single molecule analysis:** We prepared raft bearing membranes with non-fluorescent long and short rods. These membranes were doped with a very low volume fraction of fluorescent short rods (~ 0.003%) enabling us to visualize the motion of single rods within the membrane background and rafts. The membranes were simultaneously visualized in phase contrast and fluorescence microscopy to determine the number of raft-bound labeled rods, $N_f$. Standard analysis shows that the decay of the autocorrelation function of $N_f$ is given by: $\langle N_f(t) N_f(t+\Delta t) \rangle - \langle N_f(t) \rangle^2 = \left( \frac{k_{on} c_{BG}}{k_{off}} \right) e^{-k_{off} t}$ [37]. Therefore, $k_{off}$ and $k_{on}$ can be determined uniquely using this expression. These measurements were performed for equilibrium sized rafts only.

**Measurement of membrane-mediated repulsive raft interactions:** To measure effective raft-raft interactions, we used the blinking optical trap technique originally developed for measuring effective potentials of conventional spherical colloids[24]. Being enriched in shorter *fd*-Y21M rods, rafts are repelled from an optical trap. To manipulate such rafts, we create an optical plow-like configuration consisting of multiple time-shared optical traps generated by an acousto-optic deflector. Using a pair of optical plows we bring together two rafts into close proximity of each other (Supplementary Movie 5). Once the traps are switched off, the repulsive membrane-mediated interactions drive the rafts apart. It is possible to measure such interactions by repeating the blinking experiment dozens of times and analyzing the time evolution of subsequent rafts trajectories. Focusing an optical trap onto a membrane induces local distortions that could potentially affect the measurement of raft interactions. This makes it advantageous to measure interactions using blinking optical traps. In this technique, the traps are only used to bring two rafts into an initial low-probability configuration, but the actual measurement is performed when the optical traps are switched off. Upon switching the laser off, the trap-induced membrane distortions relax on a millisecond time scale, which is essentially instantaneous when compared to timescale of minutes over which two repulsive rafts drift apart.

Optical plow trap configurations were generated by time sharing a laser beam (4W, 1064 nm, Compass 1064, Coherent) using a pair of orthogonally oriented paratellurite acousto-optic deflectors (Intra-Action). The laser beam is projected onto the back focal plane of an oil-immersion objective (1.3 PlanFluor, 100X, NA 1.3) and focused onto the imaging plane. The multiple trap locations were specified using a custom LABVIEW software. Raft separations were measured as a function of time, once the traps were switched off, using standard video tracking methods[38]. The time lapse between successive frames was 100ms and the exposure time was 50ms.

The discretized probability of a raft pair being separated by distance $r_j$ at time $t+\Delta t$ is given by:

$$\rho(r_j) = \sum_i P(r_i \rightarrow r_j)\rho(r_i) \quad ,$$

where $P$ is the transition probability for a raft pair initially separated by distance $r_i$ at time $t$ to be separated by distance $r_j$ at time $t+\Delta t$ later. $\rho(r_i)$ is the probability of a raft pair being separated by a distance $r_i$ at time $t$[22]. Experimentally, $P$ is determined by binning the trajectories according to the initial and final separations in each time step. The steady state solution to Eq. 1 is equal to the equilibrium probability, $\rho(r)$, computed by calculating the eigenvector of the transition probability matrix. The inter-raft potential is given by: $U(r) = -k_B T \ln(\rho(r))$.